# Using time dependent citation rates (sales curves) for comparing scientific impacts

*Werner Marx, Hermann Schier, Ole Krogh Andersen*

*Max Planck Institute for Solid State Research, Stuttgart (Germany)*

**Abstract**

As a simple means for comparing and - if possible - predicting scientific impacts of different researchers working in the same field, we suggest comparing their "sales curves". A sales curve is the number of citations of the researcher's papers per year, the citation rate, considered as a function of time. As examples, we present the citation histories of 10 older well-cited scientists working in the same field. The sales curve is found to be highly individual, that is, there is a large variation between different scientists' sales curves. For each well-cited scientist, however, the sales curve is steadily increasing as long as he is young and active, and its slope, the citation acceleration, contains the essential information about his impact. The slope averaged over the time of activity of the scientist is roughly independent of time and is a fairly age-independent measure of his scientific impact. In physics and chemistry, well-cited active scientists have time-averaged citation accelerations at the order of 10 citations per year$^2$ or more. The normal citation acceleration is an order of magnitude smaller. We also show the sales curves for three large research institutes whose sizes have been fairly constant over the last 35 years. These sales curves are quite linear and have slopes at the order of 1 citation per scientist per year$^2$.

**Introduction**

Evaluation of research is necessary, because in a society with limited human and economic resources, priorities must be set and decisions taken. The task is difficult, however, because evaluations usually take place before consensus about the quality and significance of the research has been reached in the scientific community. Therefore, the judgment of informed experts must be relied upon, and it is important that they have access to statistical material prepared in a useful way. The present paper suggests a standard for this.

The number of scientific articles published by a researcher is readily counted, but it tells little about the importance of the work. The number of *citations*, *c*, of the published articles measures the *impact* of that work, although not necessarily its *quality*. Nowadays, citation data are often used for research evaluation, but reliable determination and competent interpretation of such data require experience and awareness.

Recently, Jorge Hirsch introduced a citation index measuring the cumulative impact of a scientist's work [1]. This so-called *h*-index has become very popular, partly because it is easily obtained provided one has access to the *Web of Science* (WoS), the search platform offered by *Thomson Scientific* [2], the former ISI, *Institute for Scientific Information*. The WoS includes the *Science Citation Index* (SCI) and has probably become the most versatile and user friendly citation analysis tool. The *h*- index is defined as the number of articles (published in SCI source journals) that have had *h* citations or more. For example, a scientist with an *h*-index of 40 will have published 40 articles, each of which has received 40 citations or more. The index can be determined by searching a given author's name under the WoS *General Search* mode and sorting the selected articles by *Times Cited* using the WoS sort command. The result is unambiguous, provided that there are either no highly cited namesakes or they can be easily removed.

The *h*-index reflects a scientist's contribution based on a broad body of publications rather than based on a few high impact articles. This suppresses the impact of a single or a few highly cited papers, sometimes being methodological contributions or reviews. The *h*-index favors scientists who consistently produce influential papers. The strong demand for quantitative and easy-to-check benchmarks supports the use of the *h*-index.

The number of citations accumulated until at present, $c(t)$, and the current value of the Hirsch-index, $h(t)$, both increase with age and are therefore not directly suitable for ranking scientists of different age. Instead, a time derivative, $c'(t)$ or $h'(t)$, might be more useful.

The *h*-index is mathematically a more complicated quantity than the number of citations; in particular is the *h*-index for a group of scientists not equal to the sum of the *h*-indices for the individuals. For that reason, the *h*-index seems unsuitable for rating research groups.

During the last decade, we have been developing search tools and have been trying out strategies for research evaluation. A task often met, was to select among candidates of age 35-50 for directing a research group or for a professorship. Another task was to rank research groups. The most useful strategy, to be described in the present paper, has been in extensive use by us and others for several years now, and it was made generally available in the WoS update end of November 2006. This strategy consists of obtaining and analyzing the time dependence of the rate of citations, $c'(t)$. These so-called "sales curves" are also suggestive for predicting the future impact of a scientist or a research group.

**Sales curves and citation acceleration**

The "Citation Report" included in the most recent WoS update enables any subscriber to obtain sales curves for any named scientist. The only difference between the WoS sales curves and the ones obtained by us [3] and discussed below is that we used the number of citing papers rather than the number of citations [4].

The time-dependent impact of a specific author, his citation rate, $c'(t)$, is thus the number of scientific articles which in a given year cites at least one of his papers. The citation rate considered as a function of time, *t*, may be seen as his sales curve.

For a single paper, the citation rate starts up slowly and the first citations appear from between months and years after the publication of the paper itself. The length of this time lag depends on the research area, the type of research, and the contents of the paper. The citation rate then peaks, typically after a few years, and finally decays slowly. Most papers receive only a few citations, but those papers which are decisive for the sales curve usually keep being cited for a long time, often decades.

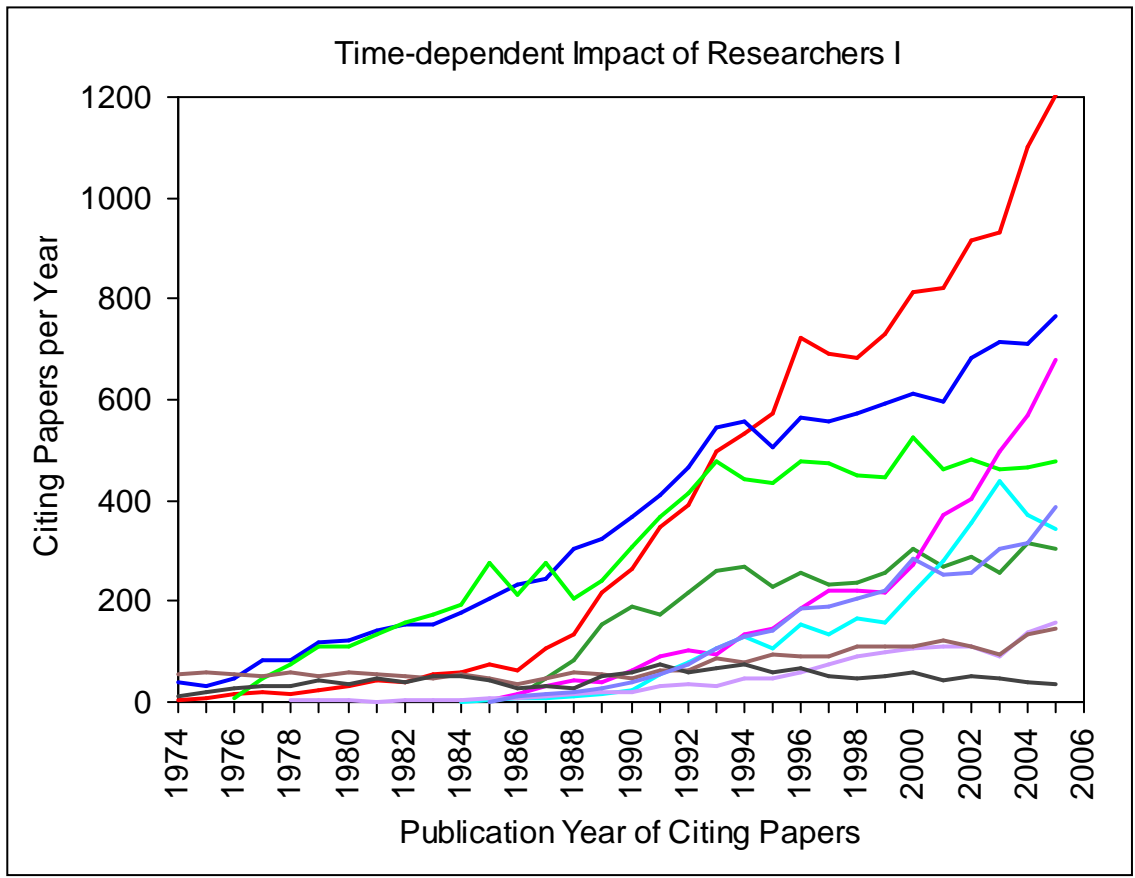

**Figure 1:** Time-dependent citation rates, or sales curves, $c'(t)$ vs. $t$, of well-cited scientists active in the same field. The citation rate is expressed as the number of citing papers per year. The slope, $c''(t)$, of the citation rate is the citation acceleration and its time average is a fairly age-independent measure of the researchers impact. The sales curves of Violet, Brown, and Black are typical for the majority of recognized researchers in this field.

Figure 1 shows the time-dependent citation rates (sales curves) of individual scientists working in the same field of research. In order to have good statistics and long histories, we have chosen well-cited scientists of age 45-65. Since the task we see as most important is to evaluate candidates of age 35-50 for leading positions or large research grants, our reason for choosing older scientists is that we want to use their histories as a guide for extrapolating the sales curves of the younger candidates. These scientists are in some sense ideals, also because their sales curves have not yet started to decline, as is the case for many scientists above age 50.

The start up of a sales curve is determined by how long it takes for the first papers to get cited and by how long it takes for the scientist to find "himself". But after he is established, his sales curve seems to rise more or less *linearly* for a long time. The linear rise of the citation rate during this period is presumably caused by the facts that, at the same time as the scientist's previous papers are being cited, he produces new papers, which get cited at an increasing rate because he is becoming more and more known. Moreover, as the scientist gets established, he may produce papers at an increasing rate, often because he has more and more people to work with.

As a fairly *age-independent* measure of a scientist's impact, we shall therefore take the slope of the sales curve, that is, the time derivative of the citation rate, $c'(t)$, i.e. the *citation acceleration*, $a(t) \equiv c''(t)$, and average it over a suitable time interval:

$$\frac{1}{t_2-t_1}\int_{t_1}^{t_2} a(t)dt = \frac{c'(t_2)-c'(t_1)}{t_2-t_1}.$$

This is the slope of the straight line connecting two points on the sales curve, i.e. of the chord. Here and in the following, $c(t)$ is the total number of citations received until time $t$. If the scientist's impact must be condensed into *one*, simple-to-evaluate number, we suggest the citation acceleration averaged over the entire career of the scientist, i.e. from the time $t_0$ of his first publication, at which point $c'(t_0) = c(t_0) = 0$, and until now, $t$:

$$\bar{a} \equiv \frac{c'(t)}{t-t_0} \approx \frac{2c(t)}{(t-t_0)^2} \equiv <a>. \quad (1)$$

So in order to evaluate the average citation acceleration, one needs to know the length of the career, $t-t_0$, and the latest citation rate (for $\bar{a}$), or the total number of citations (for $<a>$). The two measures, $\bar{a}$ and $<a>$, are equal if $c(t) = ½(t-t_0)c'(t)$, that is, if the area under the sales curve equals the area of the triangle with edge-lengths $t-t_0$ and $c'(t)$; this is for instance the case when the sales curve is a straight line. The difference $\bar{a} - <a>$ is a measure of the upwards curvature of the sales curve. If sales curves were straight lines, as is roughly the case (Figure 1), the number of citations would increase quadratically with the length of the career.

In order to provide some feeling for sales curves and in order to emphasize that citation histories are individual, we now describe those shown in Figure 1. This we do in order of decreasing total number of citations which is roughly the order of decreasing *h*-index and age (see Table I):

*Red* did not have a particularly strong impact during the first 15 years of her career. At the end of this period, she was receiving about 60 citing papers per year, that is, her citation acceleration, averaged over the first 15 years, was 4 citing papers per year$^2$. She then made a seminal contribution which strongly accelerated her citations, and this acceleration was kept up during the subsequent 20 years by citations to an increasing number of important papers produced by her and her increasing number of excellent collaborators. Currently, she is receiving over 1200 citing papers per year. This means that over the last twenty years, her average citation acceleration was 60 citing papers per year$^2$, an order of magnitude larger than before. She has opened a new, interdisciplinary direction of research and has become its established leader. Her Hirsch index is 71, 12 100 papers have cited her works, and averaged over her entire career, her citation-acceleration measures are respectively $\bar{a}$=38 and $<a>$=21 citing papers per year$^2$. The latter numbers are not only large but also show that her sales curve bends upwards.

| Scientist $t_0$ | $t-t_0$ (years) | $h$ | $c$ | citations/ citing papers | $\bar{a}$ (citing papers per year$^2$) | $<a>$ (citing papers per year$^2$) |
|---|---|---|---|---|---|---|
| Red 1971 | 34 | 71 | 12100 | 1.64 | 38 | 21 |
| Blue 1968 | 37 | 60 | 12500 | 1.45 | 20 | 18 |
| Green 1972 | 33 | 56 | 9700 | 1.48 | 16 | 18 |
| Magenta 1985 | 20 | 48 | 4400 | 1.44 | 31 | 22 |
| Spruce 1985 | 20 | 33 | 4400 | 1.35 | 14 | 22 |
| Bluish 1985 | 20 | 32 | 3200 | 1.29 | 20 | 16 |
| Cyan 1984 | 21 | 35 | 3100 | 1.38 | 15 | 14 |
| Brown 1968 | 37 | 27 | 2600 | 1.42 | 4 | 4 |
| Black 1967 | 38 | 24 | 1500 | 1.24 | 1 | 2 |
| Violet 1978 | 27 | 26 | 1300 | 1.38 | 6 | 4 |

**Table 1:** Scientist and time ($t_0$) of first publication, length ($t-t_0$) of scientific activity until 2005, Hirsch-index ($h$), number of citing papers until December 2005 ($c$), ratio of the number of citations to that of citing papers, and the average citation accelerations ($\bar{a}$ and $<a>$) defined in equation (1).

*Blue* is older and has a much smoother citation history; his citation rate is fairly linear. During the first twenty-five years of his career, the citation rate increased with a slightly increasing acceleration, peaking at about 70 citing papers per year$^2$, but then it decreased for a while, and recently picked up a bit again. He is currently receiving 700 citing papers per year, his Hirsch index is 60, and he has had 12 500 papers citing his works, yielding career-averaged citation accelerations of respectively $\bar{a}$=20 and $<a>$=18.

*Green* is slightly younger than Red and started out more successfully with a highly cited thesis work and subsequent well-cited papers. After about ten years, he made further important contributions, this time in a somewhat different field. This accelerated his citation rate to a level of almost 500 citing papers per year, where it has remained for a decade now. He is a theorist working with only a few people and can therefore hardly sustain a further increase of his citation rate, unless he moves into an exceptionally fruitful field. Unlike Red and Blue, he has not developed a popular method, the users of which provide a significant "basis" for his citation rate. His Hirsch index is 56, and with 9700 citing papers he belongs to ISI's 250 most cited physicists. The career-averaged citation accelerations are respectively $\bar{a}$=16 and $<a>$=18, with the former being smaller than the latter.

*Magenta* is much younger. She is an experimentalist whose sales curve has taken off in the last ten years, during which she has been highly successful and have also built up research groups. The slope of her sales curve is currently $a(t)$=100 citing papers per year$^2$. In 2005 she received almost 700 citations, her Hirsch index is 48, she has received a total of 4400 citing papers, and her career-averaged citation accelerations are respectively $\bar{a}$=31 and $<a>$=22, with the former exceeding the latter.

*Spruce* is of the same age as Magenta. His thesis and subsequent work, carried out in an experimental surrounding, was timely and had a large impact. But being a theorist working mostly alone on a most difficult, important problem, his citation rate has increased only slowly during the last 15 years. He has become an authority in his field and recently received a large, national prize. Currently he is receiving a little less than 300 citing papers a year, and has received 4400 citing papers in total. His Hirsch index 33, and his career-averaged citation accelerations are respectively $\bar{a}=14$ and $<a>=22$, with the latter exceeding the former.

*Bluish* is of the same age as Magenta and Spruce. He is a theorist with a nearly linear sales curve: $\bar{a}=20$ and $<a>=16$ citing papers per year$^2$. He recently received a prestigious international prize for developing a new method.

*Cyan* is a year older and spent her first five active years in the former Soviet Union. That delayed the citations to her early works. During the last 15 years, she has worked mostly on the theory of novel superconducting and magnetic materials. In 2001 and the following years, she played a key role in clearing up the mechanism behind an important, unexpected experimental discovery. This made her citation acceleration peak at 75 citing papers per year$^2$. Her Hirsch index is 35, she has received a total of 3100 citing papers, and her career-averaged citation accelerations are respectively $\bar{a}=15$ and $<a>=14$ citing papers per year$^2$.

*Brown*, *Black*, and *Violet* have sales curves typical for the majority of recognized researchers in this field with career-averaged citation accelerations between 1 and 6 citing papers per year$^2$. Since they are senior, they have substantial numbers of citations and respectable Hirsch indices, of about 25.

We have seen that citation histories are quite individual. Moreover, citation rates vary from field to field. Even within the field considered here, moving towards nano- and bio-sciences as done by Red and Magenta, does in general give more citations.

For the purpose of ranking scientists of age 35-50, the slope of the sales curve, the citation acceleration, is an appropriate measure. That would make the order of Table 1 not from top to bottom, but would move Magenta followed by Bluish and Spruce, upwards.

Now, despite individual differences, researchers with average accelerations exceeding ~10 citing papers per year$^2$ or ~15 citations per year$^2$ (in the field considered) are excellent, and the first task of an evaluation is simply to identify such candidates. The second task may be to seek desired individualities, and for that purpose, our citation histories might provide insight:

We shift all the sales curves in time such that all careers start at the same time, $t_0 \equiv 0$, and show the result in Figure 2. It is now obvious that presented with our scientists after merely 10 years of activity and using the sales-curves tool, Spruce is the only definitely outstanding young researcher, and supporting him would of course be an excellent choice. In the second rank come Green, Magenta, and Bluish, which with the benefit of hindsight is also correct. In the third rank come Cyan, Blue, and Brown, in that order. The only great error is of course that Red would not be selected.

After 15-20 years, a large gap stars to develop between the high-acceleration and the intermediate-acceleration researchers, with the gap lying around 10 citing papers per year$^2$. After 15 years, Magenta takes the lead over Spruce, Cyan, Green, and Bluish. Looking at the acceleration, rather than the rate of the citations, Red can now also be identified as promising. Finally, after 20 years, Red comes out, together with Magenta. The reasons for this are - besides talent - structural and topical.

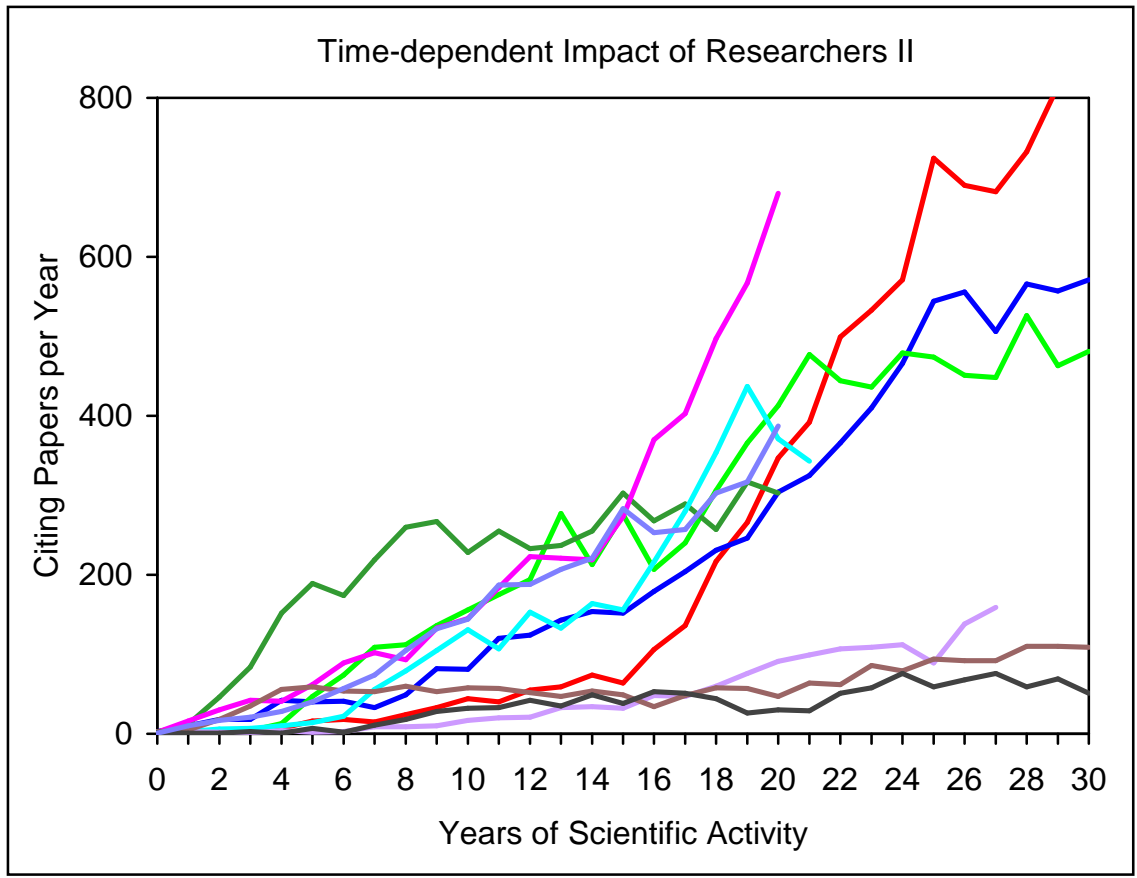

**Figure 2:** The sales curves from Figure 1, but plotted vs. the career lengths ($t$-$t_0$) with the times ($t_0$) of the first publication given in Table I.

This analysis in terms of sales curves may be extended to *research groups*. In Figure 3 we show the sales curves for three large, well-recognized institutes working in fields closely related to the one of the individual researchers considered above. The sizes of these institutes have been roughly constant over the last 25 years and the sales curves have been normalized with the number of scientists (~250 PhD students, Post docs, and staff members). These sales curves are seen to be far more linear than those of the individual researchers considered above. The differences in citation rates are mainly related to the different ages of the institutes, but, here again, the *slopes* are significant indicators of the research impact. The corresponding numbers are: $\bar{a}$ = 1.1, 0.8, and 0.5 citing papers per scientist per year$^2$. Considering the fact that each publication from such an institute has on the average 3 authors from that institute, this means that the average acceleration for an individual researcher in this field is about 2 citing papers per year[2] or 3 citations per year[2]. This fits well with the numbers given in Table 1 for Violet, Brown, and Black.

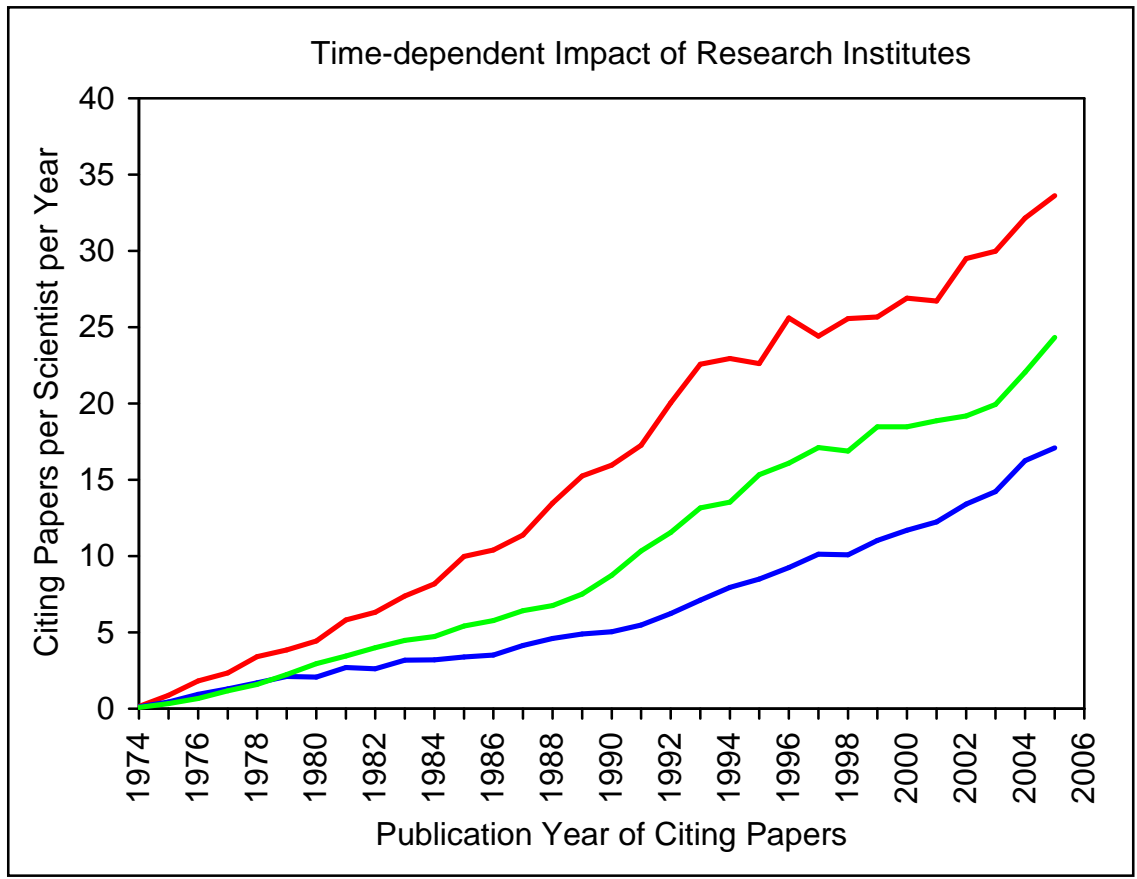

Figure 3. Sales curves for three well-recognized institutes working in related fields.

**Summary**

We have presented time-resolved citation analyses, in particular the so-called sales curves, which we have found useful for evaluating the scientific impact of a researcher or a research group, and for extrapolating into the future. The increase in the rate of cumulated citations, the citation acceleration, is a fairly age-insensitive measure of the impact. This method is suitable in particular for comparing individual scientists or research groups of widely different ages, but within the same research discipline.

**Acknowledgement**

Encouragement and useful suggestions by Klaus Kern are gratefully acknowledged.

[3] We used the search options available under the host *STN International* (URL: http://www.stn-international.de/), in particular the functions for carrying out statistical investigations. STN enables access to the SCI (file *SCIsearch*) under its own specific search system. A citation search can provide the number of *citing papers* of a specific article, or of the ensemble of articles of a specific author. The publication years of the citing papers are easily selected and their number plotted as a function of time. Self-citations may be included or easily excluded from the body of citing papers.

[4] One citing paper may comprise more than one citation of a specific author, on the average between 1.3 and 2.0 depending on the author. In general, the more papers an author publishes on related topics, the more does his number of citations exceed his number of citing papers.

Corresponding author:
Dr. Werner Marx
Max Planck Institute for Solid State Research
Heisenbergstraße 1, D-70569 Stuttgart, Germany
E-mail: w.marx@fkf.mpg.de